\documentclass[aps,twocolumn,superscriptaddress]{revtex4-2}%
\usepackage{amsfonts}
\usepackage{amsmath}
\usepackage{amssymb}
\usepackage{mathtools}
\usepackage[colorlinks=true,linkcolor=blue,citecolor=red,plainpages=false,pdfpagelabels]%
{hyperref}
\usepackage{graphicx}%
\usepackage[section]{placeins}
\usepackage{tikz}
\usetikzlibrary{quantikz}
\usepackage{amssymb}
\usepackage{physics}
\usepackage{fancyhdr}
\providecommand{\U}[1]{\protect\rule{.1in}{.1in}}
\allowdisplaybreaks

\begin{document}
\preprint{ }
\title[TITLE]{Practical Approach to Extending Baselines of Telescopes using Continuous-Variable Quantum Information}
\author{Bran Purvis}
\affiliation{Hearne Institute for Theoretical Physics and Department of Physics and Astronomy, Louisiana State University, Baton
Rouge, Louisiana 70803, USA}
\email[AFRL.RDSS.OrgMailbox@us.af.mil \\
Approved for public release; distribution is unlimited. Public Affairs release approval AFRL-2024-1118.]{}

\author{Randy Lafler}
\affiliation{Air Force Research Laboratory, Directed Energy Directorate, Kirtland AFB, NM, United States}


\author{R. Nicholas Lanning}
\affiliation{Air Force Research Laboratory, Directed Energy Directorate, Kirtland AFB, NM, United States}
\email[AFRL.RDSS.OrgMailbox@us.af.mil \\
Approved for public release; distribution is unlimited. Public Affairs release approval AFRL-2024-1118.]{}

\begin{abstract}
Interferometric telescopes are instrumental for the imaging of distant astronomical bodies, but optical loss heavily restricts how far telescopes in an array can be placed from one another, leading to a bottleneck 
in the resolution that can be achieved.
An entanglement-assisted approach to this problem has been proposed by Gottesman, Jennewein, and Croke (GJC12) [Physical Review Letters, 109(7):070503, July 2011], as a possible solution to the issue of optical loss if the entangled state can be distributed across long distances by employing a quantum repeater network.
In this paper, we propose an alternative entanglement-assisted scheme that interferes a two-mode squeezed vacuum state with the astronomical state and then measures the resulting state by means of homodyne detection.
We use a continuous-variable approach 
and compute the Fisher information with respect to the mutual coherence of the astronomical source.
We show that when the Fisher information is observed cumulatively 
at the rate at which successful measurements can be performed, our proposed scheme does not outperforms the traditional direct detection approach or the entanglement-assisted approach of GJC12.

\end{abstract}
\date{\today}
\maketitle

\pagestyle{fancy}
\cfoot{Approved for public release; distribution is unlimited. Public Affairs release approval AFRL-2024-1118.}
\lhead{}
\chead{}
\rhead{\thepage}






\textit{Introduction}---Long-baseline interferometry (LBI) is a standard technique in astronomy which is used to perform high angular resolution measurements of weak-thermal-light sources \cite{Monnier2003}.
In LBI, the optical field from an astronomical object is collected at two spatially separated telescopes then mixed to determine the mutual coherence, and thereby spatial properties of the object via the van Cittert-Zernike theorem \cite{mandel1995optical}.
LBI entered the arena of quantum-information theory when Gottesman et al. proposed a scheme, hereafter referred to as the GJC12 scheme, using shared entanglement to increase the separation between telescopes \cite{Gottesman2011}.
This is of potential interest to astronomers since the resolution of the interferometer scales as $\lambda/L$ where $\lambda$ is the wavelength and $L$ is the length of the baseline.

In the GJC12 scheme, a photon from an astronomical object couples into two telescopes, but rather than being brought to a central location to interfere with itself, an ancillary single photon is sent to the two telescopes where interference measurements are made.
The goal of this scheme is to circumvent the loss that would be observed as the baseline of the interferometer is increased, since at least in principle, the rate of the ancillary photons can be increased whereas the rate of astronomical photons cannot. 
The physics underpinning this scheme can be interpreted as a quantum teleportation; the path entanglement of the ancillary single photon is used to teleport the astronomical state at one telescope, to the other telescope, without incurring any transmission loss or decoherence.

Shortly following the GJC12 proposal, Tsang investigated the Fisher information for different measurement schemes which can be divided into two categories \cite{Tsang2011}.
One scheme, referred to as \textit{nonlocal}, involves bringing the telescope modes together or sharing entanglement between the two telescope sites.
The other scheme, referred to as \textit{local}, involves measuring the two telescope modes separately before combining the results with classical communication.
Tsang concludes that \textit{any} local measurement scheme must be significantly inferior to a nonlocal one for the estimation of the mutual coherence, at least for the single-shot measurement case.
In response, there have been investigations including imperfect resources \cite{santra2017higher, santra2020entanglement}, networks of quantum memories \cite{khabiboulline2019optical, czupryniak2021optimal}, controlled phase gates \cite{czupryniak2021optimal}, noiseless linear amplification \cite{Yang2015}, and proof-of-principle demonstrations \cite{diaz2021emulating, brown2022long, brown2021interferometry}.

In this letter, we propose a continuous-variable framework for the entanglement-assisted LBI technique.
We show that two-mode squeezed vacuum (TMSV) can be used as the entangled resource state to outperform the conventional nonlocal measurement scheme when considering the cumulative Fisher information per unit time.
Although the conventional nonlocal scheme from Ref.~\cite{Tsang2011} and the entanglement-assisted scheme from Ref.~\cite{Gottesman2011} have better single-shot scaling, the advantage is lost once photon arrival statistics are included.
Furthermore, in the limit of large mean photon number of the TMSV resource state, the cumulative Fisher information surpasses the conventional nonlocal scheme in Ref.~\cite{Tsang2011} by a factor of two.

In practice there are many competing phenomenon which can tip the scales for one scheme with respect to another.
For example, issues related to different methods of beam combining and bandwidth considerations can lend an advantage to homodyne or direct detection \cite{lawson2000principles}. 
However, in the context of large-scale quantum networking with advanced resources, our CV scheme might offer an advantage over other schemes proposed for entanglement-assisted LBI.



\begin{figure}
    \includegraphics[width=\linewidth]{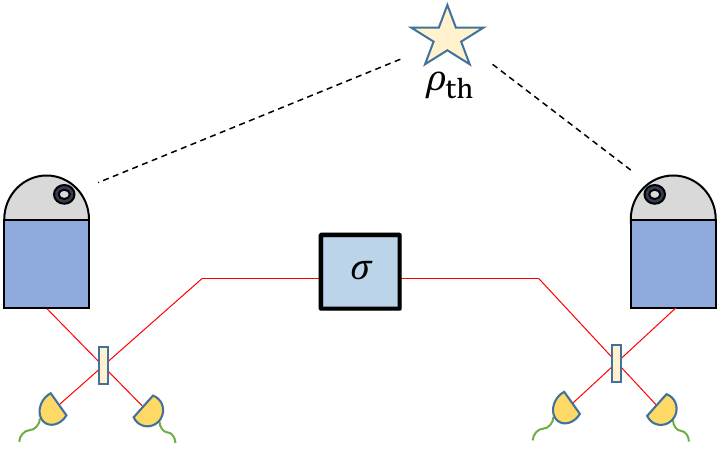}
    \caption{This diagram shows our theoretical setup, with two telescopes receiving information from the astronomical source~$\rho$. The state $\rho$ is treated as a two-mode state and mixed on two beam splitters with the TMSV state $\sigma$.}
    \label{fig:SetupDiagram}
\end{figure}

\textit{Preliminaries}---In order to model this problem, we employ a continuous-variable (CV) approach.  
Here we model the entangled state created when one mode of the astronomical state is mixed with one mode of the entangled state at each telescope, as depicted in  Figure~\ref{fig:SetupDiagram}. 

We model the astronomical state as bipartite thermal light, having a density operator of the following form: 
\begin{multline}\label{eq:rho}\rho = \frac{1}{\pi^2\text{Det}\widetilde{V}_{\rho}} \times\\
\int_{\mathbb{C}} \exp(-
	\begin{pmatrix} \alpha^* & \beta^*\end{pmatrix}
	\widetilde{V}_{\rho}^{-1}
    \begin{pmatrix} \alpha \\ \beta
    \end{pmatrix})
    \ket{\alpha , \beta}\!\!\bra{\alpha , \beta}\dd[2]{\alpha} \dd[2]{\beta},
    \end{multline}
where the covariance matrix $\widetilde{V}_{\rho}$ is given by
    \begin{equation}\label{eq:Vrho}\widetilde{V}_{\rho} \coloneqq  \frac{\epsilon}{2}
    \begin{bmatrix} 
    1 & g \\ 
    g^* & 1\\
    \end{bmatrix},
    \end{equation}
 $\epsilon$ represents the mean photon flux per coherence time, and $g = g_1 + i g_2$ represents the complex valued mutual coherence, satisfying  $|g| \leq 1$ \cite{Tsang2011}.
 
Another factor that is important to consider is the spectral bandwidth $\Delta \nu$.
In practice, $\Delta \nu$ must be severely restricted so that the effective temporal mode $\Delta t \approx 1/\Delta \nu$ is compatible with the resolution of single photon detectors as shown in \cite{deng2019quantum}.
For a photon-flux spectral density $S(\nu)$, one finds that $\epsilon = S(\nu_0) \Delta \nu \, \Delta t \approx S(\nu_0)$, and although $\epsilon$ is likely small at optical frequencies $\nu_0$, we do not make a first order approximation as done in \cite{Tsang2011}, but instead model the full state $\rho$ as above.

Ultimately, we model our system in terms of the individual position- and momentum-quadrature measurements associated with each mode, and so we introduce the following conversion between the complex amplitudes ($\alpha$ and $\beta$) and the various position- and momentum-quadrature variables corresponding to each ($x_{a,b}$ and $p_{a,b}$):
\begin{equation}\label{eq:alphabeta-xp}\begin{split}
    \alpha &= u_a + i v_a = \frac{1}{\sqrt{2}}(x_a + i p_a),\\
    \beta &= u_b + i v_b = \frac{1}{\sqrt{2}}(x_b + i p_b).\end{split}\end{equation}
In this convention, the covariance matrix of $\rho$ is as follows: 
\begin{equation}\label{eq:Vrho-xp}V_{\rho} = \epsilon\begin{bmatrix} 
	1+\epsilon^{-1} & 0 & g_1 & - g_2\\
	0 & 1+\epsilon^{-1} & g_2 & g_1\\
	g_1 & g_2 & 1+\epsilon^{-1} & 0\\
	- g_2 & g_1 & 0 & 1+\epsilon^{-1}\\
	\end{bmatrix},
	\end{equation}
where the ordering of entries is according to $\left(x_{A_1}, p_{A_1}, x_{B_1}, p_{B_1}\right)$.
The derivation of this covariance matrix can be found in Appendix~\ref{app:astro-calc}.

The covariance matrix of the TMSV state $\sigma$ is known and in this convention is given by
\begin{equation}\label{eq:Vsigma-xp}V_{\sigma} = \begin{bmatrix} 
	(2\bar{n} + 1)I_2 & 2\sqrt{\bar{n}(\bar{n} + 1)}R_{z x}(\theta) \\
	2\sqrt{\bar{n}(\bar{n} + 1)}R_{z x}(\theta) & (2\bar{n} + 1)I_2\\
	\end{bmatrix},
	\end{equation}
	where 
\begin{equation}\label{eq:Rzx}
R_{z x}(\theta) = \begin{bmatrix} 
	\cos{\theta} & \sin{\theta} \\
	\sin{\theta} & -\cos{\theta}\\
	\end{bmatrix} = \cos{\theta}\sigma_z + \sin{\theta}\sigma_x.
	\end{equation}
A more detailed calculation of this covariance matrix can be found in Appendix~\ref{app:TMSV-calc}. 


Now that we have obtained covariance matrices for both $\rho$ and $\sigma$, we can calculate the covariance matrix of the state that would result from the beam splitter mixtures that are shown in Figure~\ref{fig:SetupDiagram}.
Thus, we can determine the associated probability distribution for a given set of position- and momentum-quadrature variables by taking the reduced matrix corresponding to those positions and momenta that we wish to measure (the quadrature variables we choose to measure in this case are $x_{A_1}$, $p_{A_2}$, $x_{B_1}$, and $p_{B_2}$).
The resulting reduced covariance matrix is given below:
\begin{equation}\label{eq:final_Vr}V_{r} = \frac{1}{2}\begin{bmatrix}
    a + b & 0 & c + d & e + f\\ 
    0 & a + b & -e + f & c - d\\
    c + d & -e + f & a + b & 0\\
    e + f & c - d & 0 & a + b\\
    \end{bmatrix},
    \end{equation}
where $a = \epsilon + 1$, $b = 2\bar{n} + 1$, $c = \epsilon g_1$, $d = 2\cos{\theta}\sqrt{\bar{n}(\bar{n} + 1)}$, $e = \epsilon g_2$,  $f = 2\sin{\theta}\sqrt{\bar{n}(\bar{n} + 1)}$, and the ordering of the elements of this covariance matrix is {$x_{A_1}$, $p_{A_2}$, $x_{B_1}$, $p_{B_2}$}.
A more detailed derivation of the final state's covariance matrix and the associated probability distribution is given in Appendix~\ref{app:final-state-calcs}


\textit{Evaluating the Fisher Information}---Having obtained the reduced covariance matrix representing our desired probability distribution, our next step is to compute the Fisher information matrix of this distribution, defined as 
 \begin{equation}
 \label{eq:fisher_info}
 F(\phi) \coloneqq \int_{\rm I\!R^4}f(\phi_1, \phi_2) P(\vec{x}_r, \phi)\,d\vec{x}_r \end{equation} where  \begin{equation}
 \label{eq:fisher_info_matrix}
 f(\phi_1, \phi_2) \coloneqq \begin{bmatrix}
    \left(\frac{\partial\ln{(P)}}{\partial \phi_1}\right)^2 & \left(\frac{\partial\ln{(P)}}{\partial \phi_1}\right)\left(\frac{\partial\ln{(P)}}{\partial \phi_2}\right)\\
    \left(\frac{\partial\ln{(P)}}{\partial \phi_2}\right)\left(\frac{\partial\ln{(P)}}{\partial \phi_1}\right) & \left(\frac{\partial\ln{(P)}}{\partial \phi_2}\right)^2 \\
    \end{bmatrix},
\end{equation}
$P(\vec{x}_r, \phi)$ (written in Eq. \ref{eq:fisher_info_matrix} as P) is the probability distribution for a set of position- and momentum quadrature variables, $\vec{x}_r$. The probability distribution also depends on $\phi$, which is a complex parameter that has $\phi_1$ and $\phi_2$ as its real and imaginary parts, respectively.

We employ the Fisher information here because its inverse provides a lower bound on the uncertainty of estimating a desired unknown, which in our case is the complex coherence $g$. This Fisher information lower bound on the error is well known as the Cramer--Rao bound \cite{Rao45,Cram46}. 
In the limits of small $\bar{n}$ and large $\bar{n}$, we obtain the Fisher information matrices given in Appendix \ref{app:fisher-info-calcs}. 
In what follows, we employ the shorthands $F(g,0)$ and $F(g,\infty)$ to refer to these respective quantities.

If we were to take $\epsilon \ll 1$ as assumed in \cite{Tsang2011}, these matrices can simplify even further, which gives us
\begin{equation}\label{eq:fisher_info_small_ep}
\begin{split}
    F(g,0) &= \frac{\epsilon^2}{2}
    \begin{bmatrix}
        1  & \frac{1}{4} \\
        \frac{1}{4} & 1
    \end{bmatrix} + O(\epsilon^3), \\
    F(g,\infty) &= \epsilon^2
    \begin{bmatrix}
        1  &  1 \\
        1  &  1
    \end{bmatrix} + O(\epsilon^3),
    \end{split}
\end{equation}
where there is no $g$ dependence until terms of order $O(\epsilon^4)$ and beyond.
Therefore $F(g,\infty)$ will be larger than $F(g,0)$ within the desired range of $\epsilon$.
In other words, increasing the squeezing will reduce the lower Cramer–Rao bound on measurements of $g$, an intuitive result that is nevertheless critical to convey.

To elaborate on the meaning of these results, it is worth explaining what the Fisher information we calculated represents. For example, $F(g,0)$  corresponds to mixing the astronomical state with a vacuum state. $F(g,\infty)$, on the other hand, corresponds to the case of  homodyne measurements taken at both detectors. In other words, as $\bar{n}$ increases towards $+\infty$, the Fisher information approaches the result of taking the Fisher information in the  homodyning case.


\textit{Cumulative Fisher Information}---In this section we investigate the cumulative Fisher information over $M$ measurements $F^{(M)} = MF$ \cite{Tsang2011}.
Here, we are careful to derive the number of successful measurements that are possible in each case, bearing in mind that each of the schemes have different spectral bandwidths $\Delta \nu$ in practice. 
For example, the Navy Precision Optical Interferometer (NPOI) takes the broadband astronomical photons and performs simultaneous measurements over 16$+$ spectral channels, each of which is approximately 15 THz \cite{ garcia2016vision, van2018many}.
In contrast, homodyne measurements are typically limited to the GHz range due to the limited bandwidth of the local oscillator.
Similarly, the GJC12 scheme has GHz spectral filtering requirements due to the limited detector resolution, which is tens of picoseconds in practice \cite{deng2019quantum}. 
This disparity will lead to a drastic difference in the number of ``successful'' measurements.
Since the astronomical state is defined per coherence time, these bandwidths dictate that the conventional scheme of the NPOI would receive approximately $15\mathrm{THz}/1\mathrm{GHz}=15,000$ more copies of the state per second per channel, and thus accommodate more measurements $M$. 

However, we prefer to show impartiality by setting the bandwidths equal for all the schemes.
In the context of global-scale quantum networking, some of the constraints pose relatively straightforward optical-engineering challenges. 
For example, there are methods for lifting the narrow-bandwidth limitation on optical homodyne detection \cite{shaked2018lifting}. 
Similarly, one can envision a spectrally-multiplexed GJC12 scheme in which a broadband ancillary photon is sent to interfere with the astronomical photon and split into multiple spectral bands before detection.
For example, one could use a dispersive element to spread the state spectrally over a high-resolution single-photon-resolving camera.
Thus, for the remainder of this section we will assume the spectral bandwidth $\Delta \nu$ is the same for all the schemes.

First, we consider homodyne detection. 
As is discussed qualitatively in Ref.~\cite{Tsang2011}, the downside of homodyne detection is that it cannot distinguish when no photon is arriving at the receiver to provide information about the unknown parameters, and thus must include vacuum fluctuations as potentially useful measurements. 
In other words, homodyne is an unconditional measurement, comprised of a photo-current difference regardless of outcome, and we assume
\begin{equation}\label{eq:homo}
M_h \approx \Delta \nu.
\end{equation}

An alternate measurement scheme consists of performing conditional measurements based on photon counting. 
For example, in Ref.~\cite{Tsang2011}, local and nonlocal detection schemes are investigated by assuming $\epsilon \ll 1$, approximating the state $\rho$ to first order, and using a POVM based on single photon detection events to calculate the single-shot Fisher information.
This technique reduces the number of successful measurements due to the arrival statistics of single photons, but the unsuccessful measurements can be identified, and thus, discarded.
Therefore, the number of successful counts observed would be 
\begin{equation}\label{eq:count}
M_{\mathrm{count}} \approx \Delta \nu ,
\end{equation}
which is the same as what we get for homodyne measurements.
For the GJC12 scheme, a photon from the astronomical source must arrive at the same time as a photon from the EPR source, but at the opposite telescope from where the EPR photon arrives. 
If both photons arrive at the same telescope, the measurement is discarded. This leads to a large number of unsuccessful measurements, but much like in the previous case, these cases are identifiable and can be discarded.
This leads to the result that
\begin{equation}\label{eq:Gott}
M_{\mathrm{G}} \approx \Delta \nu
\end{equation}
which is once again the same result seen in the previous cases. 

Now, we can find expressions for the norm of the cumulative Fisher informations for five particularly interesting cases. 
First, we use the Fisher information from Eqs.~\eqref{eq:F-} and \eqref{eq:F+} and find
\begin{equation}\label{eq:CVTN}
\Vert F_{\mathrm{CV}}(g,\infty)\Vert_1  \geq 2 \, \epsilon^2 , \qquad 
\Vert F_{\mathrm{CV}}(g,0)\Vert_1 \geq  \epsilon^2, 
\end{equation}
where $\Vert F \Vert_1$ is the trace norm and only lowest order terms have been retained.
We will also use the results from Ref.~\cite{Tsang2011}
\begin{equation}\label{eq:TsangResults}
\Vert F_{\mathrm{DD}}\Vert_1 \geq \epsilon , \qquad 
\Vert F_{\mathrm{L}}\Vert_1 \geq  \epsilon^2, \qquad
\Vert F_{\mathrm{G}}\Vert_1 \geq  \frac{\epsilon}{2},
\end{equation}
where the subscripts DD, L, and G stand for direct detection, local, and GJC12 schemes, respectively. 
Combining Eqs.~\eqref{eq:homo}--\eqref{eq:TsangResults}, we find the cumulative Fisher information for each of the schemes:
\begin{equation}\label{eq:Cumulative}
\begin{split}
\Vert F_{\mathrm{CV}}^{(M)}(g,\infty)\Vert_1 & \geq 2 \, \Delta \nu \, \epsilon^2, \\
\Vert F_{\mathrm{CV}}^{(M)}(g,0)\Vert_1 &\geq \Delta \nu \, \epsilon^2, \\
\Vert F_{\mathrm{DD}}^{(M)}(g)\Vert_1 &\geq \Delta \nu \, \epsilon, \\
\Vert F_{\mathrm{L}}^{(M)}(g)\Vert_1 &\geq \Delta \nu \, \epsilon^2 , \\ 
\Vert F_{\mathrm{G}}^{(M)}(g)\Vert_1 &\geq \dfrac{1}{2} \, \Delta \nu \, \epsilon,
\end{split}
\end{equation}
where $M_h$ is used for the CV scheme, $M_{\mathrm{count}}$ is used for direct detection and local schemes, $M_{\mathrm{G}}$ is used for the GJC12 scheme, and only lowest order terms in $\epsilon$ have been retained.
This is an interesting result because our entanglement-based scheme would be expected to perform similarly to that of the theoretical performance of the GJC12 scheme, but instead it only performs slightly better than local measurements would.
This suggests that, despite our CV scheme being entanglement-based, the homodyne detections that we perform in it would be the most significant limiting factor in the quality of our Fisher information.
\begin{figure}[] 
 \includegraphics[width=1\columnwidth]{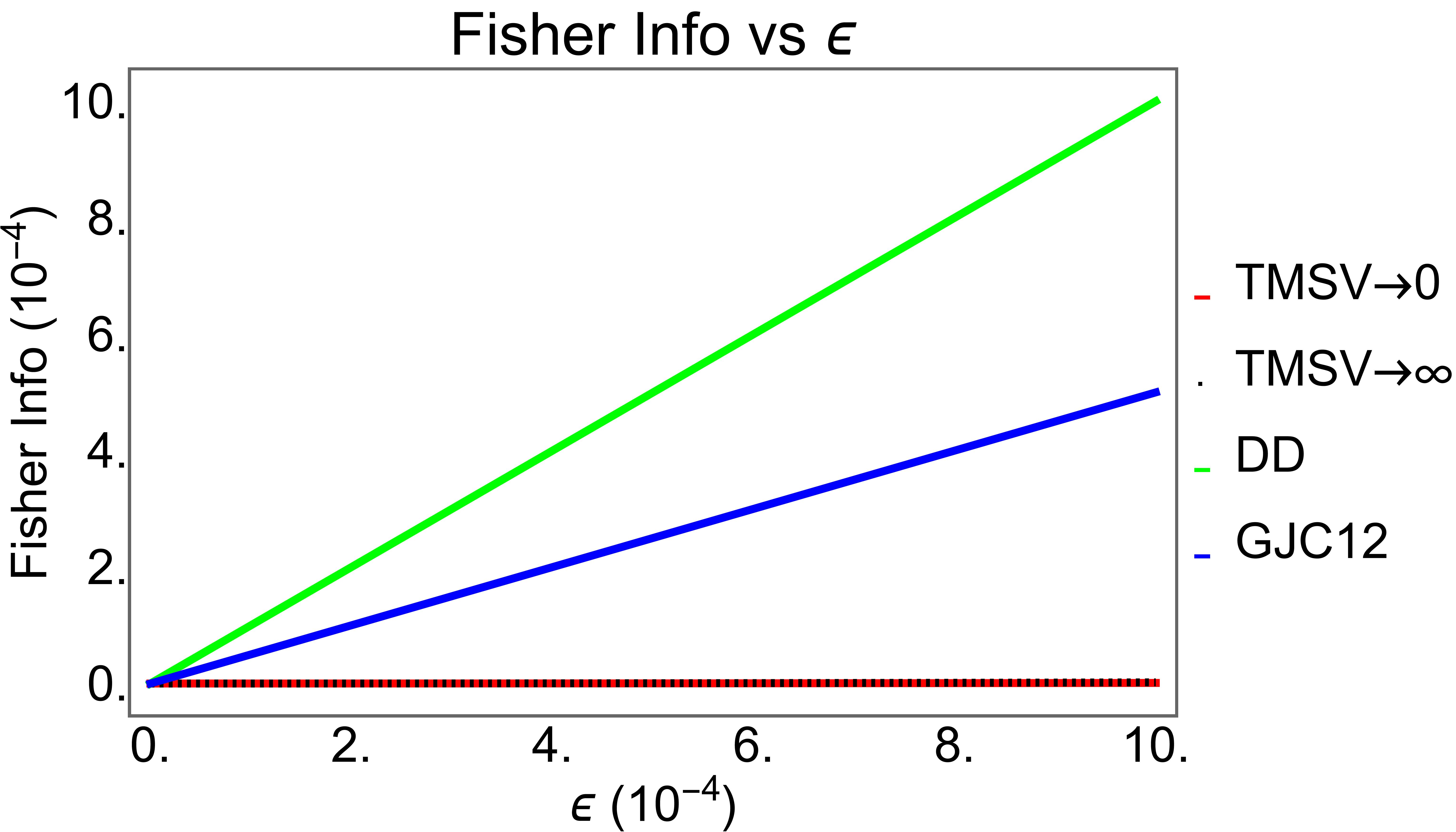}
 \caption{The cumulative Fisher information from Eq.~\ref{eq:Cumulative} for the CV when $\bar{n} \rightarrow 0$ (red), CV when $\bar{n} \rightarrow \infty$ (dashed black), direct detection (green), and GJC12 (blue) schemes. $\epsilon$ remains the average photon flux per coherence time.
 }
 \label{fig:CumulativeInfoAll}
\end{figure}

In Fig.~\ref{fig:CumulativeInfoAll} we ignore the common factor $\Delta \nu$ and plot the Fisher information from Eq.~\eqref{eq:Cumulative}.
One can see the difference in scaling between our proposed scheme and direct detection, but the GJC12 scheme's performance shows that entanglement-based schemes like our proposed one have a lot of unrealized potential.

\textit{Conclusion}---In this paper, we showed that, in its current iteration, our entanglement-assisted framework measures the error of the mutual coherence, $g$, at a worse degree of fidelity than what would be achieved with other non-local methods like direct detection.
Despite this, we believe that this discrepancy has more to do with our choice of taking homodyne measurements rather than an underlying issue with entanglement-assisted framework in general.
By taking homodyne detecton, vacuum noise in the system can affect the fidelity of measurements, and unlike with direct detection, it is more difficult to simply remove the unimportant data and make the most of the collected information.
Also, it is important to note that the theoretical limit on entanglement-assisted schemes is much higher than what our scheme can currently achieve, which definitely suggests that improvements are possible.
With some modifications, the disadvantage our proposed scheme currently displays could potentially be eliminated, which would be an incredible result. 
When that day comes, an entanglement-assisted framework would finally hold the advantage originally chased by \cite{Gottesman2011}. 
Namely, an opportunity to create arrays of telescopes with longer baselines. 


The aforementioned opportunity appears for this framework due to the reduced rates of photon loss that come about when there is an intermediary entangled state. 
Such a state will shorten the distance that photons would have to travel in order to reach locations where relevant measurements would be taken. 
In addition to that, with further advances in quantum repeater technology, the entangled state could be easily distributed to multiple locations across wide distances so that this framework could be implemented for arrays with large numbers of telescopes as well.

Another advantage of our framework comes from the CV approach we chose. It allowed us to model the astronomical and entangled states as Gaussian states with easily calculable covariance matrices.
With that, formulae for the Fisher information of these states is also easily obtained, which gave us a suitable metric for comparing our framework to other suggested detection methods.
As such, making modifications to the framework that could result in higher fidelity is something that should be easily possible.

In the future, there are many directions to pursue from here.
First, we can analyze how other forms of noise that would exist in realistic implementations of this setup might affect our projected lower bound on error.
Most significantly, experimenting with how the fidelity improves when different detection methods are used could lead to more promising results.
We can also consider how our method could be implemented across quantum repeater networks, and how those networks might work if implemented.
One could also say that the results of this paper could be applied to other regimes of $\epsilon$, rather than only the limit where it is considered to be very small.
Ref.~\cite{Tsang2011}  covered several such regimes, and it would be interesting to revisit some of those calculations with the results of our paper in mind.


\begin{acknowledgments}
The authors thank Dr. Mark M. Wilde for his helpful discussions.

The views expressed are those of the authors and do not reflect the official guidance or position of the United States Government, the Department of Defense or of the United States Air Force.
The appearance of external hyperlinks does not constitute endorsement by the United States Department of Defense (DoD) of the linked websites, or the information, products, or services contained therein. The DoD does not exercise any editorial, security, or other control over the information you may find at these locations.

Approved for public release; distribution is unlimited. Public Affairs release approval AFRL-2024-1118.
\end{acknowledgments}

\bibliographystyle{unsrt}
\bibliography{citations_copy}

\appendix

\section{Calculation of the Covariance Matrix of the Astronomical State~$\rho$}
\label{app:astro-calc}

Tsang defines $V_{\rho}$ as
\begin{equation}\label{eq:TsangRho}V_{\rho} = \begin{bmatrix} 
	\langle a^{\dagger}a\rangle & \langle b^{\dagger}a\rangle \\
	\langle a^{\dagger}b\rangle & \langle b^{\dagger}b\rangle\\
	\end{bmatrix},
	\end{equation} where $a = \frac{x_a + ip_a}{\sqrt{2}}$ and $b = \frac{x_b + ip_b}{\sqrt{2}}$ \cite{Tsang2011}.
	
We want $V_{\rho}$ in an (x, p) notation rather than the ($\alpha$, $\beta$) notation Tsang et al writes it in. Working with the above relation, we get
\begin{equation}\label{eq:XPDiag}
    \begin{split}\langle a^{\dagger}a\rangle & =\left\langle\frac{(x_a - ip_a)(x_a + ip_a)}{2}\right\rangle\\
    \langle a^{\dagger}a\rangle & = \frac{1}{2}\langle x_a^2 + p_a^2 + i(x_a p_a - p_a x_a)\rangle.
    \end{split}\end{equation}
To simplify this expression, we use the fact that $\langle (x_a p_a - p_a x_a)\rangle = i$ (we take $\hbar = 1$), and thus we obtain
\begin{equation}\label{eq:XPDiagSimp}
    \langle a^{\dagger}a\rangle = \frac{1}{2}(\langle x_a^2\rangle + \langle p_a^2\rangle -1).
    \end{equation}

Now, we use the assumption used in Tsang that \begin{equation}\label{eq:TsangAssumeEps}\langle a^{\dagger}a\rangle = \frac{\epsilon}{2},
\end{equation}

and set our right hand side equal to $\frac{\epsilon}{2}$, to end up with
\begin{equation}\label{eq:XPDiagEps}\begin{split}
    \langle x_a^2\rangle  + \langle p_a^2\rangle - 1 & = \epsilon\\
    \langle x_a^2\rangle  + \langle p_a^2\rangle  & = \epsilon+1.
\end{split}\end{equation}
	
We assume that $\langle x_a^2\rangle $ and $\langle p_a^2\rangle $ each get an equal share of the right hand side (these measurements are analogous to measuring photon number), so,
\begin{equation}\label{eq:XPDiagElements}
\langle x_a^2\rangle  = \langle p_a^2\rangle = \frac{\epsilon + 1}{2}.
\end{equation}
Similarly, $x_b^2$ and $p_b^2$ will be equal to the same quantity, which is trivial to prove through the same method of calculation.
	
For the off-diagonal elements, we have
\begin{equation}\label{eq:OffDiagFirst}\begin{split}
    \langle b^{\dagger}a\rangle & = \left\langle\frac{(x_b - ip_b)(x_a + ip_a)}{2}\right\rangle \\
    \langle b^{\dagger}a\rangle & = \frac{1}{2}\langle x_b x_a + p_b p_a + i(x_b p_a - p_b x_a)\rangle,
\end{split}\end{equation} and
\begin{equation}\label{eq:OffDiagSecond}\begin{split}
    \langle a^{\dagger}b\rangle & = \left\langle\frac{(x_a - ip_a)(x_b + ip_b)}{2}\right\rangle \\
    \langle a^{\dagger}b\rangle & = \frac{1}{2}\langle x_a x_b + p_a p_b + i(x_a p_b - p_a x_b)\rangle.
    \end{split}\end{equation}
	
As we can see, $\langle b^{\dagger}a\rangle $ and $\langle a^{\dagger}b\rangle $ are complex conjugates of one another, as we should expect. Tsang sets 
\begin{equation}\label{eq:TsangAssumeOffDiag}\begin{split}
    \langle b^{\dagger}a\rangle  & = \frac{\epsilon g}{2}\\
    \langle a^{\dagger}b\rangle  & = \frac{\epsilon g^*}{2},
\end{split}\end{equation} where g is a complex number such that $g = g_1 + ig_2$. We will refer to $g_1$ as Re[g] and $g_2$ as Im[g] for the time being, just to make their relation to g more clear.
	
In order to obtain a useful result from this, we will have to add or subtract the two off-diagonal elements.
I now write them once more with some simplification as
\begin{equation}\label{eq:XPOffDiagEps}\begin{split}
    \langle x_b x_a + p_b p_a + i(x_b p_a - p_b x_a)\rangle & = \epsilon g\\
    \langle x_a x_b + p_a p_b + i(x_a p_b - p_a x_b)\rangle & =\epsilon g^*.
\end{split}\end{equation}
Summing them and dividing both sides by 2 gives
\begin{equation}\label{eq:XPOffDiagSum}
    \langle x_a x_b\rangle +\langle p_a p_b\rangle = \epsilon g_1,
    \end{equation}
while subtracting the latter from the former and dividing by 2i gives \begin{equation}\label{eq:XPOffDiagSubtract}
    \langle x_b p_a\rangle - \langle p_b x_a\rangle = \epsilon g_2,
\end{equation}
where $g_1$ is the real part of g and $g_2$ is the imaginary part like before.
	
Once again, these quantities will split evenly between both relevant expectation values, but this proof is not trivial and so I will show it below.
In order to prove this, we need to take these expectation values manually by tracing them over the state $\rho$. $\rho$ is given as follows after simplifying the form given in Tsang's letter \cite{Tsang2011} and normalizing it: 
\begin{multline}\label{eq:TsangRhoIntegral}\rho =
\\
\frac{1}{\pi^2\text{Det}V_{\rho}}\int_{\mathbb{C}} e^{-
	\begin{pmatrix} \alpha^* & \beta^*\end{pmatrix}
	V_{\rho}^{\text{-}1}
    \begin{pmatrix} \alpha \\ \beta
    \end{pmatrix}}
    \ket{\alpha , \beta}\bra{\alpha , \beta}\dd[2]{\alpha} \dd[2]{\beta},
    \end{multline}
where we let 
\begin{equation}\label{eq:TsangRhoCovMatrix}V_{\rho} = \frac{\epsilon}{2}\begin{bmatrix} 
    1 & g \\ 
    g^* & 1\\
    \end{bmatrix}.
    \end{equation}
The trace of any operator $\hat{\xi}$ on $\rho$ will hence be given by
\begin{multline}
\label{eq:TsangRhoExpectationValue}\Tr{\hat{\xi}\rho} =
\\ \frac{4}{\pi^2\epsilon^2 (1 - \abs{g}^2)} 
    \int_{\mathbb{C}}  f(\alpha , \beta) \bra{\alpha , \beta}\hat{\xi}\ket{\alpha , \beta}\dd[2]{\alpha} \dd[2]{\beta},
    \end{multline}
where
\begin{equation}\label{eq:TsangRhoExpectationValueExponentialPart}
    f(\alpha , \beta) = \exp\left\lbrack-2\frac{\abs{\alpha}^2 + \abs{\beta}^2 - \alpha^* \beta g - \alpha \beta^* g^*}{\epsilon - \epsilon \abs{g}^2}\right\rbrack.
\end{equation}
    
All of the operators we wish to find the trace of are the products of an operator that only acts on the $\alpha$ mode and another operator that only acts on the $\beta$ mode so 
\begin{equation}\label{eq:ExpectationSplit}\bra{\alpha , \beta} \hat{\xi}_a \hat{\xi}_b \ket{\alpha , \beta} = \bra{\alpha}\hat{\xi}_a \ket{\alpha}\bra{\beta} \hat{\xi}_b  \ket{\beta} .
\end{equation}
The only $\hat{\xi}_{a,b}$ operators we will be working with, however, are $\hat{x}_{a,b}$ and $\hat{p}_{a,b}$, so
\begin{equation}\label{eq:ExpectationSplitSimp}\bra{\alpha}\hat{\xi}_a \ket{\alpha}\bra{\beta} \hat{\xi}_b \ket{\beta} = \xi_a \xi_b
.
\end{equation}
    
This gives us enough information to use a change of variables that will provide a nice intermediary between $\alpha , \beta$ notation and $x, p$ notation.
We define variables $u_{a,b}$ and $v_{a,b}$ such that
\begin{equation}
\label{eq:AlphaBetToUV}\begin{split} \alpha = u_a + i v_a &= \frac{1}{\sqrt{2}}(x_a + i p_a),\\
\beta = u_b + i v_b &= \frac{1}{\sqrt{2}}(x_b + i p_b).\end{split}\end{equation}
Defined this way, we can use the fact that $\dd[2]{\alpha} = \dd{\text{Re}\lbrack \alpha\rbrack}\dd{\text{Im}\lbrack \alpha\rbrack}$ (and similarly for $\beta$) to convert our contour integral over complex space into a traditional area integral integrating with respect to $\dd{u_{a,b}}\dd{v_{a,b}}$ as well as make use of the fact that
\begin{equation}\label{eq:UVToXP}\begin{split}x_{a,b} &= \sqrt{2}u_{a,b},\\
p_{a,b} &= \sqrt{2}v_{a,b},\end{split}
\end{equation}
assuming that $x$ and $p$ are both real numbers.
    
Hence, we have
\begin{equation}\label{eq:TsangIntegralUV}\langle \hat{\xi}_a \hat{\xi}_b\rangle = \frac{8}{\pi^2\epsilon^2 (1 - \abs{g}^2)} 
    \int_{-\infty}^{\infty}\int_{-\infty}^{\infty} \gamma_a\gamma_b e^{-2\Phi (u,v)} \dd{^2u}\dd{^2v}\end{equation}
where
\begin{multline}
\label{eq:PhiUV}\Phi (u,v) =
\\
\frac{\abs{\vec{s}}^2 - 2g_1 (u_a u_b + v_a v_b) + 2g_2 (u_a v_b - v_a u_b)}{\epsilon - \epsilon \abs{g}^2},
\end{multline}
\begin{equation}\label{eq:VecS}
    \abs{\vec{s}}^2 = u_a^2 + u_b^2 + v_a^2 + v_b^2,
\end{equation}
and
\begin{equation}\label{eq:GammaAB}
    \gamma_{a,b} = u_{a,b} \text{ or } v_{a,b}.
\end{equation}
    
Integrating for each case, we get
    
\begin{equation}\label{eq:OffDiagElements}\begin{split}
\langle \hat{x}_a \hat{x}_b\rangle &= \frac{\epsilon}{2}g_1, \\
\langle \hat{p}_a \hat{p}_b\rangle &= \frac{\epsilon}{2}g_1 ,\\
\langle \hat{x}_b \hat{p}_a\rangle &= \frac{\epsilon}{2}g_2 ,\\
\langle \hat{p}_b \hat{x}_a\rangle &= - \frac{\epsilon}{2}g_2, \end{split}
\end{equation}
which is exactly what we set out to prove.
	
\section{Calculation of the Covariance Matrix of the Two-Mode Squeezed Vacuum}

\label{app:TMSV-calc}

The purpose of this section is to prove that
\begin{equation}\label{eq:SigmaCovResult}V_{\sigma} = \begin{bmatrix} 
	(2\bar{n} + 1)I_2 & 2\sqrt{\bar{n}(\bar{n} + 1)}R_{z x}(\theta) \\
	2\sqrt{\bar{n}(\bar{n} + 1)}R_{z x}(\theta) & (2\bar{n} + 1)I_2\\
	\end{bmatrix},
\end{equation}
where $R_{z x}$ is defined by \ref{eq:Rzx}.
We start with the result obtained in Gerry and Knight for the Two-Mode Squeeze Operator \cite{GerryKnight2005}:
\begin{equation}\label{eq:GerryKnightTMSOp}
    \hat{S}_2(\xi) = \exp{\xi^*\hat{a}\hat{b}-\xi\hat{a}^{\dagger{}}\hat{b}^{\dagger{}}},
\end{equation}
which is for a value of $\xi = r e^{i\theta}$.
However, it must be noted that this Squeezing operator is given as the time evolution operator of the Hamiltonian, so, despite being unitary, it is not quite the same as the symplectic unitary used in Serafini's notation.
Essentially, the only real difference is that the time evolution operator depends on $e^{-i\hat{H}}$ wheras Serafini's notation uses $e^{i\hat{H}}$, so if we multiply the exponent by $-i$, we can use Serafini's methods for determining the covariance matrix of the state.
So, we want to make use of the formula 
	
\begin{equation}\label{eq:SerafiniHamiltonianOp}\hat{H} = \frac{1}{2}\hat{r}^\text{T}H\hat{r}\end{equation}
from Serafini \cite{Serafini2017}. 
In order to find H, we convert $-i\hat{H}$ into a form that depends on $\hat{r}$:
\begin{multline}\label{eq:SerafiniHamiltonianOpEq}
    -i \hat{H} =
    \\
    -\frac{i}{2}\lbrack(\xi^*-\xi)(\hat{x}_a\hat{x}_b - \hat{p}_a\hat{p}_b)+i(\xi^*+\xi)(\hat{x}_a\hat{p}_b + \hat{p}_a\hat{x}_b)\rbrack .
\end{multline}
Using $\xi^* - \xi = -2i r\sin{\theta}$ and $\xi^* + \xi = 2 r\cos{\theta}$, we get
\begin{multline}\label{eq:SerafiniHamiltonianOpEqSimp}
    -i\hat{H} =
    \\
    \frac{1}{2}\lbrack-2 r\sin{\theta}(\hat{x}_a\hat{x}_b - \hat{p}_a\hat{p}_b) + 2 r\cos{\theta}(\hat{x}_a\hat{p}_b + \hat{p}_a\hat{x}_b)\rbrack
\end{multline}.
    
Conveniently, this can be expressed in matrix form as
\begin{equation}\label{eq:SerafiniHamiltonianOpEqMatrix}-i\hat{H} = \frac{1}{2}\hat{r}^\text{T}
    \begin{bmatrix} 
	0 & 0 & -r\sin{\theta} & r\cos{\theta} \\
	0 & 0 & r\cos{\theta} & r\sin{\theta} \\
	-r\sin{\theta} & r\cos{\theta} & 0 & 0 \\
	r\cos{\theta} & r\sin{\theta} & 0 & 0 \\
	\end{bmatrix}\hat{r}\end{equation}
where the matrix in the center of this equation is $H r$.
Putting this in terms of the symplectic matrix $\omega$, we get
\begin{equation}\label{eq:SerafiniHamiltonianOpEqSimpler}
\Omega H = \begin{bmatrix} 
	0 & 0 & r\cos{\theta} & r\sin{\theta} \\
	0 & 0 & r\sin{\theta} & -r\cos{\theta} \\
	r\cos{\theta} & r\sin{\theta} & 0 & 0 \\
	r\sin{\theta} & -r\cos{\theta} & 0 & 0 \\
	\end{bmatrix}\end{equation}
which is a form we desire because of Serafini's assertion that the covariance matrix will be given as $V_{\sigma} = S I_4 S^{\text{T}} = S S^{\text{T}}$ where $S = e^{\Omega H r}$ \cite{9781482246346}. 
Conveniently, $S^{\text{T}} = S$, so $S S^{\text{T}} = S S$. But let us first determine the form of S.
S is given as
\begin{multline}\label{eq:SerafiniSEq}S = e^{\Omega H r} = \\
    \begin{bmatrix} 
	\cosh{r} & 0 & \cos{\theta}\sinh{r} & \sin{\theta}\sinh{r} \\
	0 & \cosh{r} & \sin{\theta}\sinh{r} & -\cos{\theta}\sinh{r} \\
	\cos{\theta}\sinh{r} & \sin{\theta}\sinh{r} & \cosh{r} & 0 \\
	\sin{\theta}\sinh{r} & -\cos{\theta}\sinh{r} & 0 & \cosh{r} \\
	\end{bmatrix},
\end{multline} 
which will allow us to find $V_{\sigma}$, given as
\begin{equation}\label{eq:SerafiniVEqSimp}V_{\sigma} = 
    \begin{bmatrix} 
	(2\bar{n} + 1)I_2 & 2\sqrt{\bar{n}(\bar{n} + 1)}R_{z x}(\theta) \\
	2\sqrt{\bar{n}(\bar{n} + 1)}R_{z x}(\theta) & (2\bar{n} + 1)I_2\\
	\end{bmatrix},
\end{equation}
which uses the definition 
\begin{equation}
    \label{eq:nBarToR}2\bar{n}+1 \coloneqq \cosh{2r},
\end{equation}
as well as the matrix given in \ref{eq:Rzx}.
	
\section{Calculation of the Covariance Matrix of the Resultant Mixed State and Its Associated Probability Distribution}

\label{app:final-state-calcs}

In order to represent the mixing of $\rho$ and $\sigma$ on beam splitters, we will model the modes of each such that the product state between them for the case of a two-telescope array can be given as
\begin{equation}\label{eq:RhoSigmaProductState}
    \rho_{A_{1}B_{1}} \otimes \sigma_{A_{2}B_{2}}.
\end{equation}
Using this notation, there will be a beam splitter that mixes the modes $A_1$ and $A_2$ and a beam splitter than mixes the modes $B_1$ and $B_2$, as shown by \begin{equation}\label{eq:RhoSigmaBSMixEq}
    (R_{A_{1}A_{2}} \otimes R_{B_{1}B_{2}})^\dagger(\rho_{A_{1}B_{1}} \otimes \sigma_{A_{2}B_{2}})(R_{A_{1}A_{2}} \otimes R_{B_{1}B_{2}}),
\end{equation} where 
\begin{equation}\label{eq:BeamSplitter2Mode}
    R_{m_{1}m_{2}} = \frac{1}{\sqrt{2}}\begin{bmatrix} 
	1 & 0 & 1 & 0 \\
	0 & 1 & 0 & 1 \\
	-1 & 0 & 1 & 0 \\
	0 & -1 & 0 & 1
	\end{bmatrix}
\end{equation} and $m$ represents either mode $A$ or mode $B$.

Now, the above equations show transformations on the relevant states, but all we desire is to know how the covariance matrices will be altered by this transformation. 
Serafini's textbook \cite{Serafini2017} provides much of the intuition for how to solve such a problem.
For example, \cite{Serafini2017} shows that the covariance matrix of a product state can be easily modelled as 
\begin{equation}\label{eq:CovForProductState}
    V_{\rho} \oplus V_{\sigma},
\end{equation} and that if a quadratic unitary $\hat{U}$ acts on a Gaussian state, the state's covariance matrix $V$ changes according to
\begin{equation}\label{eq:UnitaryModifyingV}
    V \rightarrow UVU^T,
\end{equation}
where U is a transformation matrix associated with $\hat{U}$.
$R_{m_{1}m_{2}}$ can be trivially shown to be unitary, and so the stated transformation will be valid.

Let $V_f$ correspond to the covariance matrix of our final mixed state with the ordering of the elements given by {$x_{A_1}$, $p_{A_1}$, $x_{A_2}$, $p_{A_2}$, $x_{B_1}$, $p_{B_1}$, $x_{B_2}$, $p_{B_2}$}.
Given what has been shown above, 
\begin{equation}\label{eq:FinalV}
    V_f = R_2 V_{\text{product}}R_{2}^T,
\end{equation} or more explicitly,
\begin{equation}\label{FinalVMoreExplicit}
    V_f = \frac{1}{2}
    \begin{bmatrix}
        V_D & V_{12}\\
        V_{21} & V_D
    \end{bmatrix}
\end{equation} where 
\begin{align}\label{eq:VSubMatrices}
    V_D &=
    \begin{bmatrix}
        a + b & 0 & -a + b & 0\\
        0 & a + b & 0 & -a + b\\
        -a + b & 0 & a + b & 0\\
        0 & -a + b & 0 & a + b\\
    \end{bmatrix}, \\
    V_{12} &=
    \begin{bmatrix}
        c + d & -e + f & -c + d & e + f\\
        e + f & c - d & -e + f & -(c + d)\\
        -c + d & e + f & c + d & -e + f\\
        -e + f & -(c + d) & e + f & c - d\\
    \end{bmatrix}, \\
    V_{21} &=
    \begin{bmatrix}
        c + d & e + f & -c + d & -e + f\\
        -e + f & c - d & e + f & -(c + d)\\
        -c + d & -e + f & c + d & e + f\\
        e + f & -(c + d) & -e + f & c - d\\
    \end{bmatrix},
\end{align} where $a = \epsilon + 1$, $b = 2\bar{n} + 1$, $c = \epsilon g_1$, $d = 2\cos{\theta}\sqrt{\bar{n}(\bar{n} + 1)}$, $e = \epsilon g_2$, and $f = 2\sin{\theta}\sqrt{\bar{n}(\bar{n} + 1)}$.
    
Finally, we find the reduced covariance matrix corresponding to the measurements we wish to take. What this means is, given that the covariance matrix has its elements ordered according to {$x_{A_1}$, $p_{A_1}$, $x_{A_2}$, $p_{A_2}$, $x_{B_1}$, $p_{B_1}$, $x_{B_2}$, $p_{B_2}$}, and we wish to measure {$x_{A_1}$, $p_{A_2}$, $x_{B_1}$, $p_{B_2}$}, the reduced matrix will be made up of the elements {$(x_{A_1},x_{A_1})$, $(x_{A_1},p_{A_2})$, $(x_{A_1},x_{B_1})$, $(x_{A_1},p_{B_2})$, $(p_{A_2},x_{A_1})$, $(p_{A_2},p_{A_2})$, $(p_{A_2},x_{B_1})$, $(p_{A_2},p_{B_2})$, $(x_{B_1},x_{A_1})$, $(x_{B_1},p_{A_2})$, $(x_{B_1},x_{B_1})$, $(x_{B_1},p_{B_2})$, $(p_{B_2},x_{A_1})$, $(p_{B_2},p_{A_2})$, $(p_{B_2},x_{B_1})$, $(p_{B_2},p_{B_2})$} as shown in Eq. \ref{eq:final_Vr}.

In order to find the probability distribution, We use the formula:
    
\begin{equation}\label{eq:SerafiniProbDist}
    P(\vec{x}_r, g) = \frac{1}{\sqrt{(2\pi)^4 Det(V_r)}} \exp{-\vec{x}_{r}^T V_{r}^{-1} \vec{x}_r / 2}
\end{equation} given by \cite{Serafini2017}.
It is with this probability distribution that we are able to obtain the Fisher information, as shown in the next Appendix section.

\section{Calculation of the Fisher Information Taken With Respect To The Mutual Coherence}

\label{app:fisher-info-calcs}

By plugging the probability distribution obtained with the formula in Eq. \ref{eq:SerafiniProbDist} (where $V_r$ is given by Eq. \ref{eq:final_Vr}), we can make use of Eqs. \ref{eq:fisher_info} \& \ref{eq:fisher_info_matrix} to find the Fisher information:

     \begin{widetext}
     \begin{align}\label{eq:F-}
     \lim_{\bar{n}\rightarrow 0} F(g,\bar{n}) & = \left(\frac{\sqrt{2} \epsilon}{4 + 4 \epsilon - (| g |^2-1) \epsilon^2}\right)^2 \begin{bmatrix}
    4 + 4 \epsilon + (1 + g_{1}^2 - g_{2}^2) \epsilon^2 & 2 g_1 g_2 \epsilon^2\\ 
    2 g_1 g_2 \epsilon^2 & 4 + 4 \epsilon + (1 - g_{1}^2 + g_{2}^2)  \epsilon^2\\
    \end{bmatrix},\\
    \label{eq:F+}
    \lim_{\bar{n}\rightarrow \infty}
    F(g,\bar{n}) & = \left(\frac{\epsilon}{1 + 2 \epsilon - (| g |^2-1) \epsilon^2}\right)^2 \begin{bmatrix}
    1 + 2 \epsilon + (1 + g_{1}^2 - g_{2}^2) \epsilon^2 & 2 g_1 g_2 \epsilon^2\\ 
    2 g_1 g_2 \epsilon^2 & 1 + 2 \epsilon + (1 - g_{1}^2 + g_{2}^2) \epsilon^2\\
    \end{bmatrix},
    \end{align}
     \end{widetext}
     where $F(g,\bar{n})$ is the Fisher information of $g$, evaluated for a fixed mean photon number $\bar{n}$. 
If we were to take $\epsilon \ll 1$ as assumed in \cite{Tsang2011}, these matrices can simplify to the results shown in Eq. \ref{eq:fisher_info_small_ep}.

\end{document}